\input harvmac
\input graphicx
%
%
\ifx\includegraphics\UnDeFiNeD\message{(NO graphicx.tex, FIGURES WILL BE IGNORED)}
\def\figin#1{\vskip2in}
\else\message{(FIGURES WILL BE INCLUDED)}\def\figin#1{#1}
\fi
\def\Fig#1{Fig.~\the\figno\xdef#1{Fig.~\the\figno}\global\advance\figno
 by1}
%
%
%
%
\def\Ifig#1#2#3#4{
\goodbreak\midinsert
\figin{\centerline{
\includegraphics[width=#4truein]{#3}}}
\narrower\narrower\noindent{\footnotefont
{\bf #1:}  #2\par}
\endinsert
}
%
%
\font\ticp=cmcsc10

\def\hf{{1\over 2}}
\def\calo{{\cal O}}

\def\calh{{\cal H}}

\def\cald{{\cal D}}

\def\calA{{\cal A}}
\def\barh{{\bar h}}
\def\calL{{\cal L}}
\def\mp{{M_{p}}}
\def\mpm{{M_{p}^{-1}}}
\def\mpmt{{M_{p}^{-2}}}

\def\rtg{{\sqrt{-g}}}

\def\mthsu{\mathsurround=0pt  }
\def\leftrightarrowfill{$\mthsu \mathord\leftarrow\mkern-6mu\cleaders
  \hbox{$\mkern-2mu \mathord- \mkern-2mu$}\hfill
  \mkern-6mu\mathord\rightarrow$}
 \def\overleftrightarrow#1{\vbox{\ialign{##\crcr\leftrightarrowfill\crcr\noalign{\kern-1pt\nointerlineskip}$\hfil\displaystyle{#1}\hfil$\crcr}}}
\overfullrule=0pt
%
%
\lref\BanksTCP{
  T.~Banks,
  ``T C P, Quantum Gravity, The Cosmological Constant And All That..,''
  Nucl.\ Phys.\  B {\bf 249}, 332 (1985).
}
\lref\Hartlesc{
  J.~B.~Hartle,
  ``Prediction in quantum cosmology,''
in {\it   Cargese 1986, proceedings, gravitation in astrophysics}, 329-360. 
}
\lref\Kiefer{
  C.~Kiefer,
  {\sl Quantum gravity,}
  Int.\ Ser.\ Monogr.\ Phys.\  {\bf 124}, 1 (2004).
}
\lref\HartGS{
  J.~B.~Hartle,
  ``Ground State Wave Function Of Linearized Gravity,''
  Phys.\ Rev.\  D {\bf 29}, 2730 (1984).
}
\lref\Page{
  D.~N.~Page,
  ``Information in black hole radiation,''
  Phys.\ Rev.\ Lett.\  {\bf 71}, 3743 (1993)
  [arXiv:hep-th/9306083].
}
\lref\tHooholo{
  G.~'t Hooft,
  ``Dimensional reduction in quantum gravity,''
  arXiv:gr-qc/9310026.
  }
\lref\Sussholo{
  L.~Susskind,
  ``The World as a hologram,''
  J.\ Math.\ Phys.\  {\bf 36}, 6377 (1995)
  [arXiv:hep-th/9409089].
}
\lref\Mald{
  J.~M.~Maldacena,
  ``The large N limit of superconformal field theories and supergravity,''
  Adv.\ Theor.\ Math.\ Phys.\  {\bf 2}, 231 (1998)
  [Int.\ J.\ Theor.\ Phys.\  {\bf 38}, 1113 (1999)]
  [arXiv:hep-th/9711200].
}
\lref\BHcomp{
  L.~Susskind, L.~Thorlacius and J.~Uglum,
  ``The Stretched horizon and black hole complementarity,''
  Phys.\ Rev.\ D {\bf 48}, 3743 (1993)
  [arXiv:hep-th/9306069].
}
\lref\SGinfo{S.~B.~Giddings,
  ``Quantum mechanics of black holes,''
  arXiv:hep-th/9412138.
}
\lref\BHMR{
  S.~B.~Giddings,
  ``Black holes and massive remnants,''
  Phys.\ Rev.\ D {\bf 46}, 1347 (1992)
  [arXiv:hep-th/9203059].
}
\lref\Astroinfo{
  A.~Strominger,
  ``Les Houches lectures on black holes,''
  arXiv:hep-th/9501071.
}
\lref\EHNS{
  J.~R.~Ellis, J.~S.~Hagelin, D.~V.~Nanopoulos and M.~Srednicki,
  ``Search For Violations Of Quantum Mechanics,''
  Nucl.\ Phys.\  B {\bf 241}, 381 (1984).
}
\lref\Mathur{
  S.~D.~Mathur,
  ``Resolving the black hole information paradox,''
  Int.\ J.\ Mod.\ Phys.\  A {\bf 15}, 4877 (2000)
  [arXiv:gr-qc/0007011].
}
\lref\Joerev{
  J.~Polchinski,
  ``String theory and black hole complementarity,''
  arXiv:hep-th/9507094.
}
\lref\Hawkunc{
  S.~W.~Hawking,
  ``Breakdown Of Predictability In Gravitational Collapse,''
  Phys.\ Rev.\ D {\bf 14}, 2460 (1976).
}
\lref\SGIII{
  S.~B.~Giddings,
  ``(Non)perturbative gravity, nonlocality, and nice slices,''
  Phys.\ Rev.\  D {\bf 74}, 106009 (2006)
  [arXiv:hep-th/0606146].
}
\lref\GMH{
  S.~B.~Giddings, D.~Marolf and J.~B.~Hartle,
  ``Observables in effective gravity,''
  arXiv:hep-th/0512200.
}
\lref\GaGi{
  M.~Gary and S.~B.~Giddings,
  ``Relational observables in 2d quantum gravity,''
  arXiv:hep-th/0612191.
}
\lref\LoTh{D. Lowe and L. Thorlacius, ``Comments on the information problem," arXiv:hep-th/0601059.}
\lref\Hawkrec{
  S.~W.~Hawking,
  ``Information loss in black holes,''
  Phys.\ Rev.\ D {\bf 72}, 084013 (2005)
  [arXiv:hep-th/0507171].
}
\lref\LPSTU{
  D.~A.~Lowe, J.~Polchinski, L.~Susskind, L.~Thorlacius and J.~Uglum,
  ``Black hole complementarity versus locality,''
  Phys.\ Rev.\ D {\bf 52}, 6997 (1995)
  [arXiv:hep-th/9506138].
}
\lref\GiLitwo{
  S.~B.~Giddings and M.~Lippert,
  ``The information paradox and the locality bound,''
  Phys.\ Rev.\ D {\bf 69}, 124019 (2004)
  [arXiv:hep-th/0402073].
}
\lref\GiLione{
  S.~B.~Giddings and M.~Lippert,
  ``Precursors, black holes, and a locality bound,''
  Phys.\ Rev.\ D {\bf 65}, 024006 (2002)
  [arXiv:hep-th/0103231].
}
\lref\BPS{
  T.~Banks, L.~Susskind and M.~E.~Peskin,
  ``Difficulties For The Evolution Of Pure States Into Mixed States,''
  Nucl.\ Phys.\ B {\bf 244}, 125 (1984).
}
\lref\CGHS{
  C.~G.~Callan, S.~B.~Giddings, J.~A.~Harvey and A.~Strominger,
  ``Evanescent black holes,''
  Phys.\ Rev.\ D {\bf 45}, 1005 (1992)
  [arXiv:hep-th/9111056].
}
\lref\Hawkrad{
  S.~W.~Hawking,
  ``Particle Creation By Black Holes,''
  Commun.\ Math.\ Phys.\  {\bf 43}, 199 (1975)
  [Erratum-ibid.\  {\bf 46}, 206 (1976)].
}
\lref\Waldunpub{R.~M.~Wald, unpublished.}
\lref\HoMa{
  G.~T.~Horowitz and J.~Maldacena,
  ``The black hole final state,''
  JHEP {\bf 0402}, 008 (2004)
  [arXiv:hep-th/0310281].
}
\lref\SGII{
  S.~B.~Giddings,
  ``Locality in quantum gravity and string theory,''
  Phys.\ Rev.\  D {\bf 74}, 106006 (2006)
  [arXiv:hep-th/0604072].
}\lref\SGI{
  S.~B.~Giddings,
``Black hole information, unitarity, and nonlocality,''
  Phys.\ Rev.\  D {\bf 74}, 106005 (2006)
  [arXiv:hep-th/0605196].
}
\lref\SGinfo{S.~B.~Giddings,
  ``Quantum mechanics of black holes,''
  arXiv:hep-th/9412138\semi
  ``The Black hole information paradox,''
  arXiv:hep-th/9508151.
}
\lref\Astroinfo{
  A.~Strominger,
  ``Les Houches lectures on black holes,''
  arXiv:hep-th/9501071.
}
\lref\Vaid{P.C. Vaidya, Proc. Indian Acad. Sci {\bf A33}, 264 (1951).}
\lref\tHooftTQ{
  G.~'t Hooft,
  ``The scattering matrix approach for the quantum black hole: An overview,''
  Int.\ J.\ Mod.\ Phys.\ A {\bf 11}, 4623 (1996)
  [arXiv:gr-qc/9607022].
}
\lref\BaFi{T. Banks and W. Fischler, ``Space-like singularities and thermalization," hep-th/0606260.}
\lref\NAH{N. Arkani-Hamed, talk at the KITP conference {\sl String phenomenology 2006}.}
\lref\GiMa{S.B. Giddings and D. Marolf, work in progress.}
\lref\BanksCG{
  T.~Banks,
  ``Some Thoughts on the Quantum Theory of de Sitter Space,''
  arXiv:astro-ph/0305037.
}
\lref\BanksWR{
  T.~Banks, W.~Fischler and S.~Paban,
  ``Recurrent nightmares?: Measurement theory in de Sitter space,''
  JHEP {\bf 0212}, 062 (2002)
  [arXiv:hep-th/0210160].
}
\lref\Bankslittle{
T.~Banks,
``Cosmological breaking of supersymmetry or little Lambda goes back to  the future. II,''
arXiv:hep-th/0007146.
}
\lref\Fischler{W. Fischler, unpublished (2000);
W. Fischler, ``Taking de Sitter seriously," Talk given at
Role of Scaling Laws in Physics and Biology (Celebrating
the 60th Birthday of Geoffrey West), Santa Fe, Dec. 2000.}
\lref\BoussoFQ{
  R.~Bousso,
  ``Adventures in de Sitter space,''
  arXiv:hep-th/0205177.
}
\lref\DysonPF{
  L.~Dyson, M.~Kleban and L.~Susskind,
  ``Disturbing implications of a cosmological constant,''
  JHEP {\bf 0210}, 011 (2002)
  [arXiv:hep-th/0208013].
}
\lref\GoheerVF{
  N.~Goheer, M.~Kleban and L.~Susskind,
 ``The trouble with de Sitter space,''
  JHEP {\bf 0307}, 056 (2003)
  [arXiv:hep-th/0212209].
}
\lref\AhnZK{
  D.~Ahn,
  ``On the final state boundary condition of the Schwarzschild black hole,''
  Phys.\ Rev.\  D {\bf 74}, 084010 (2006)
  [arXiv:hep-th/0606028].
}
\lref\PageSC{
  D.~N.~Page,
  ``Is Black Hole Evaporation Predictable?,''
  Phys.\ Rev.\ Lett.\  {\bf 44}, 301 (1980).
}
\lref\HorowitzVP{
  G.~T.~Horowitz,
 ``Tachyon condensation and black strings,''
  JHEP {\bf 0508}, 091 (2005)
  [arXiv:hep-th/0506166].
}
\lref\HorowitzMR{
  G.~T.~Horowitz and E.~Silverstein,
  ``The inside story: Quasilocal tachyons and black holes,''
  Phys.\ Rev.\  D {\bf 73}, 064016 (2006)
  [arXiv:hep-th/0601032].
}
\Title{\vbox{\baselineskip12pt
\hbox{hep-th/0703116}
}}
{\vbox{\centerline{Quantization in black hole backgrounds}
}}
\centerline{{\ticp Steven B. Giddings}\footnote{$^\star$}{Email address: giddings@physics.ucsb.edu} 
}
\centerline{ \sl Department of Physics}
\centerline{\sl University of California}
\centerline{\sl Santa Barbara, CA 93106-9530}
\bigskip
\centerline{\bf Abstract}
Quantum field theory in a semiclassical background can be derived as an approximation to quantum gravity from a weak-coupling expansion in the inverse Planck mass.  Such an expansion is studied for evolution on ``nice-slices" in the spacetime describing a black hole of mass M.  Arguments for a breakdown of this expansion are presented,  due to significant gravitational coupling between fluctuations, which is consistent with the statement that existing calculations of information loss in black holes are not reliable.  For a given fluctuation, the coupling to subsequent fluctuations becomes of order unity by a time of order $M^3$.  Lack of a systematic derivation of the weakly-coupled/semiclassical  approximation would indicate a role for the non-perturbative dynamics of gravity, and possibly for the proposal that such dynamics has an essentially non-local quality.

\Date{}

\newsec{Introduction}

Hawking's discovery of black hole radiance\Hawkrad\ has produced a paradox that may be as important to finding a quantum description of gravity as the paradox of the classical instability of matter was in the foundation of quantum mechanics.  There is no commonly-accepted explanation for what is wrong with Hawking's original argument that black holes destroy information\Hawkunc.  This is despite widespread belief that black holes respect unitary quantum evolution, which is now shared by originator of the paradox himself\Hawkrec. 

From the beginning it has seemed  that the challenge of Hawking radiation requires a major conceptual advance.  Any proposal regarding the fate of information apparently conflicts with a cherished principle of physics.\foot{For more in-depth reviews, see \refs{\SGinfo,\Astroinfo}.}  Hawking's original proposal of information loss violates unitary quantum evolution, and closer studies of the resulting dynamics\refs{\EHNS,\BPS} argue it leads to
to drastic violations of energy conservation.  Alternatively, unitary evolution might be preserved if information escapes  after Hawking's calculation breaks down in the Planck phase of evaporation.  But  basic quantum restrictions on how rapidly information can be transmitted with a given energy then apparently imply a long-lived remnant.  Remnants seem equally disasterous; to encode the information from an arbitrarily large black hole, they are expected to come in infinite species, and correspondingly they would be infinitely produced in any process with sufficient total energy.  These outcomes suggest that  information must escape early in black hole evaporation, but Hawking's argument to the contrary apparently requires only very basic assumptions such as locality.

Attempts at a resolution of this paradox have increasingly focussed on some breakdown of locality\refs{\BHMR\tHooholo-\Sussholo}, and this possibility is also strongly suggested by aspects of string theory such as the AdS/CFT correspondence\Mald.  However, what has not been clear in much of this discussion is  the mechanism  for a breakdown of locality that obviates Hawking's argument.

The apparently fundamental nature of the paradox suggests comparison with the paradoxes that arose in the transition from classical to quantum physics, and the instability of the classical atomic model seems particularly apt.  In a gedanken history, were it not for the fact that the stability of matter is so critical, classical physicists might have reasoned that the problem lay in the singularity in classical evolution, when the electron reaches the charge center, and might have tried various methods to regulate this singularity.  History shows us this is not the correct approach; indeed, the problem is that classical physics simply ceases to be a valid description long before a classical electron reaches the nucleus.  Classical physics does not exhibit a {\it breakdown}, but rather quantum physics {\it supplants} it at scales set by the Bohr radius.  

This suggests a possible way that the black hole paradox could be resolved:  a local quantum field theory description simply ceases to be a good description in the vicinity of the Schwarzschild radius, long before it breaks down near the singularity.  Like in the case of the classical atom, we may not see the failure in the present physical framework.  What is of course missing is the more complete framework within which such a statement could be properly understood.  

Another different alternative is that local quantum field theory actually does signal its limitations through some explicit failure, following a slightly different path than that from classical  to quantum.  It is worth pursuing this possibility as far as possible, in an effort to find any relevant clues to the problem.  

This paper will investigate the question of the validity of Hawking's argument for loss of information\refs{
\Hawkrad,\Hawkunc}.  With more recent refinements, the argument goes as follows.  First, one can draw a family of smooth spacelike slices through the interior of a black hole of large mass $M$, avoiding the region of strong curvature, and asymptoting to usual time slices.  Away from the singularity, the geometry does not fluctuate, so the evolution on these slices can be described in terms of local quantum field theory on a semiclassical background.  This evolution produces states on late-time slices that contain significant quantum entanglement between interior and exterior.  Tracing over  internal states gives a density matrix that describes physics outside, and the entanglement implies that this density matrix has large entropy, of order $M^2$.  This parametrizes  missing information assumed to vanish with the black hole.

We begin by describing in general the steps that can be used to justify the approximation of quantum field theory in a semiclassical background; this arises from a weak coupling expansion in the inverse Planck mass $\mp$.  This approximation is described both from the functional and canonical perspectives, and then a more complete argument for information loss is given.   Section three then carefully examines the validity of this expansion in the specific situation of evolution on the slices described above, called ``nice slices" in \LPSTU.  Such a careful treatment reveals an apparent flaw in this weak-coupling expansion, arising from strong coupling between fluctuations at early and late times.  This effect is uniquely gravitational.  Section four then examines this effect more carefully, discussing the diagrammatic expansion, the connection to large relative boosts in the black hole geometry, possible lessons for locality, possible extensions to inflationary cosmology, and other issues.  Concluding remarks appear in section five.

This paper is thus consistent with the second possibility above, that the semiclassical approximation breaks down due to strong gravitational effects.\foot{For other potential sources of breakdown, see \refs{\PageSC\HorowitzVP-\HorowitzMR}.}  This has been previously advocated in \refs{\GiLione\GiLitwo\LoTh\SGI\SGII-\SGIII}.\foot{This issue has also been explored in unpublished work by N. Arkani-Hamed.}  If indeed there is no reliable calculation of the information destroyed by a black hole, there is no information paradox.  This paper improves on the previous arguments, accounting for the effects of  fluctuations on the nice-slice states used to calculate the density matrix and thus its entropy.  Ref.~\LPSTU\ attempted to produce a related conclusion due to effects of long strings, but failed to find an unambiguous effect\Joerev; moreover, \SGII\ argues that  in high-energy collisions (which are closely related to black hole evolution) there is no apparent evidence for the relevant such long-string effects, but rather strong gravitational dynamics plays a central role.

\newsec{Perturbative gravity and the semiclassical description}

A good starting point is a description of the quantization of gravity, reviewing its perturbative quantization either in path integral or canonical terms.  Let us particularly focus  on the origin of the semiclassical approximation.  This has been previously described in \refs{\BanksTCP\Hartlesc-\Kiefer} as derived in a WKB approach, but we will find that in a slightly different approach it can be derived in a weak-coupling expansion about a semiclassical spacetime.  We will find that this approximation indeed yields dynamics described as quantum field theory in a curved background, and thus demonstrate how Hawking's argument for information loss  arises.

\subsec{Path-integral quantization}

A concrete approach to quantization  is the path integral.  Consider a theory with gravity coupled to matter, which will for simplicity be taken to be a single massless scalar field, with lagrangian
\eqn\scal{S[g,\phi]=-\hf\int d^4x\rtg (\nabla \phi)^2\ .}
Also for simplicity we work in four-dimensions, although the generalization to higher dimensions is straightforward.
The gravitational action is taken to be 
\eqn\ehlag{S[g]={M_p^2\over 2}\int d^4x \rtg R + M_p^2 \oint \sqrt{\vert^3g\vert} K\ ,}
which includes the well-known extrinsic-curvature surface term, and where $M_p$ is the Planck mass, defined as $\mp^{-2}=8\pi G$.  

In order to have a well-defined time, we will work with such amplitudes defined over geometries that are asymptotically flat. Consider in particular a geometry described in terms of a family of spacial slices, with coordinates $x^i$, labeled by a time variable $t$.  The ADM decomposition of the metric is
\eqn\admd{ds^2 = -(N^2-N_i N^i) dt^2 + 2 N_i dt dx^i + g_{ij} dx^i dx^j}
where $N$ is the lapse, $N^i$ is the shift, and spatial (latin) indices are raised and lowered with the spatial metric $g_{ij}$ which describes the intrinsic geometry of the slices.  For asymptotically flat geometries, the asymptotic time is
\eqn\atime{T=\int dt N(\infty)\ .}

We are interested in amplitudes for transitions between slices with definite asymptotic time difference.  These can be defined as 
\eqn\fctlint{\calA[\Psi_i,\Psi_f,T] = \int_{\Psi_i}^{\Psi_f} \cald g \cald \phi e^{i S[g] + iS[g,\phi]} \delta(T-\int dt N(\infty))\ .}
In this expression, the state is expressed in terms of data on the metric and field and their momenta.  For example, in the field representation, one might specify the metric $g_{ij}(x)$ and the field $\phi(x)$ on the initial and final slices.

A standard approach is to expand the metric as a fluctuating field about some background metric,
\eqn\gfluc{g_{\mu\nu} = g_{0\mu\nu} + M_p^{-1} h_{\mu\nu}\ .}
For example, in such an expansion the ground-state wavefunction has been derived in the quadratic approximation in \refs{\HartGS}.  To do so, one rewrites \fctlint\ in terms of an integral over the perturbations $h_{\mu\nu}$, and gauge fixes.  The expansion of the action \ehlag\ in powers of $\mp^{-1}$ is of the form
\eqn\lagexp{S[g_0+h] = S[g_0]+ \int d^4x {\sqrt -g_0} \left[\mp^{-1} h\delta_gS[g_0] - \hf(Dh)^2 + M_p^{-1} h (\nabla h)^2+\cdots\right] = \sum_n S^g_n\ .}
Here $(Dh)^2$ is the kinetic term for $h$
(integration of $(Dh)^2$ by parts gives the Lichnerowicz laplacian), and this and higher terms can be straightforwardly worked out in detail.  One can likewise expand $S[g,\phi]$ in an expansion in $h$ and $\phi$.  For simplicity we assume that the classical background for $\phi$ vanishes (although this could easily be generalized), and as a result
\eqn\phiexp{S[g_0+h,\phi] = \int d^4 x \sqrt{-g_0}\left\{ -\hf(\nabla^0\phi)^2 - \hf M_p^{-1}h^{\mu\nu} T^\phi_{\mu\nu} + M_p^{-2}\calo\left[h^2(\partial \phi)^2\right]\right\}=\sum_nS_n^\phi\ ,}
where the leading term is $n=2$, to match \lagexp.

A well-defined amplitude requires gauge fixing the diffeomorphism symmetry
\eqn\diffxm{g_{\mu\nu}\rightarrow g_{\mu\nu} + M_p^{-1}\calL_\xi g_{\mu\nu} = g_{\mu\nu} + M_p^{-1}\calL_\xi g_{0\mu\nu} + M_p^{-2}\calL_\xi h_{\mu\nu}, }
whose linearized version is thus 
\eqn\lindiff{\delta h_{\mu\nu} = 2\nabla^0_{(\mu}\xi_{\nu)}\ .}
This is accomplished by applying a suitable gauge condition; one useful example at the linear level and in vauco is Hilbert-DeDonder gauge, defined via
\eqn\hbdef{\barh_{\mu\nu} = h_{\mu\nu} -\hf g_{0\mu\nu} h}
as
\eqn\hilbg{\nabla_0^{\mu} \barh_{\mu\nu}=0\ .}
Such a gauge has a residual set of symmetries, corresponding to diffeomorphisms satisfying
\eqn\resid{\nabla_0^2 \xi_\mu =0\ .}
These are fixed by also specifying conditions on $h$ on the initial/final surfaces, resulting in the expected 10-4-4=2 degrees of freedom; these slice conditions can alternately be thought of as a choice of the initial/final slices.

Denote the general such combined gauge conditions by
\eqn\gaugecond{C^I(h)=0\ ;}
these are enforced by inserting the expression
\eqn\gfmeas{\Delta[C^I(h)] det(\delta C^I)}
into the path integral, where $\Delta$ is a functional delta function, and $det(\delta C^I)$ is the corresponding Fadeev-Popov determinant.

The resulting amplitude takes the form
\eqn\pertamp{\eqalign{\calA[\Psi_i,\Psi_f,; T] = e^{iS[g_0]} \int_{\Psi_i}^{\Psi_f}& \cald_{g_0+h}h \cald_{g_0+h}\phi \Delta[C^I(h)] det(\delta C^I) \cr &e^{iM_p^{-1}\int d^4x \sqrt{-g_0} h\delta_gS[g_0]} e^{i(S_2+S_3+\cdots)}\ .}}
Here, as is explicitly indicated, the integration measures depend on the perturbation, as does the gauge-fixing measure, and  we define $S_n=S_n^g+S_n^\phi$.  The initial and final states are specified by giving wavefunctionals of the perturbations ${}^3h$ of the three-geometry  and fields on the initial/final slices.  In the case with $g_{0\mu\nu}=\eta_{\mu\nu}$, ref.~\HartGS\ computes an explicit ground-state wavefunction for the linearized theory from such an expression.

To simplify the perturbation expansion about a more general metric $g_0$, one should choose it to eliminate the linear terms in $h$.  There is both the explicit term and the terms from the measures.  A useful prescription starts from the quantum stress tensor in the initial state,
\eqn\qstress{\langle T_{\mu\nu}\rangle_i=\langle\Psi_i|T_{\mu\nu}|\Psi_i\rangle = 2i\mp{\delta\over \delta h^{\mu\nu}} ln[Z_i]\ ,}
where $Z_i$ is the functional integral \pertamp\ without the classical and linear terms, and with $T=\infty$ and $\Psi_f=\Psi_i$.  Specifically, define $g_0$ as a solution to 
\eqn\backstate{2{\delta S\over \delta g^{\mu\nu}}[g_0] =  \langle T_{\mu\nu}\rangle_i\ .}
In a black hole spacetime, such an equation precisely determines the semiclassical metric sourced by the average Hawking flux; such equations were studied explicitly in the two-dimensional context in \refs{\CGHS}.
For such a $g_0$, the linear term is eliminated, with the understanding\refs{\SGI} that at the same time we must replace $T^\phi$ in \phiexp\ by its ``normal-ordered" version,
\eqn\tno{\Delta T^\phi = T^\phi - \langle T^\phi\rangle_i\ ,}
and likewise for the corresponding quadratic expression in $\nabla h$ in \lagexp\ that corresponds to a gravitational stress pseudo-tensor.  The final result is of the form
\eqn\ampgff{\calA[\Psi_i,\Psi_f; T] = e^{iS[g_0,T]} \int_{\Psi_i}^{\Psi_f} \cald h \Delta[C^I(h)] det(\delta C^I) \cald \phi   e^{i(S_2+S_3+\cdots)}\ }
with the replacement \tno\ implicit.

The semiclassical limit now appears as the weak-coupling limit $M_p\rightarrow\infty$, or more specifically $p_i\cdot p_j\ll M_p^2$ where $p_i\cdot p_j$ is a typical momentum invariant.  In this case, one drops all terms beyond $S_2$ in \ampgff, as in \HartGS.  The corresponding expression computes quantum amplitudes for the field $\phi$ and the spin-two perturbation $h_{\mu\nu}$ in the fixed background geometry $g_0$ -- this limiting procedure precisely produces the amplitudes of quantum field theory in a curved background.  For a black hole, taking this limit holding $M/M_p^2$ fixed preserves a non-trivial geometry.

Of course, when one attempts to evaluate \ampgff\ beyond quadratic order, one finds the usual UV divergences of quantum gravity.  These are commonly believed to be {\it short-distance} issues.  One could imagine regulating these by a cutoff, and working with an effective description of gravity below the cutoff.  For example, in the context of a higher-dimensional theory, one could even think of string theory as a mechanism to provide a UV regulator.  We will thus regard \ampgff\ as a useful effective description of long-distance gravity.  A focus of this paper will be, assuming that such short distance problems can be cured, are there any unanticipated issues in the {\it macroscopic} dynamics of gravity, that can be understood from this expression?

\subsec{Hamiltonian quantization}

The hamiltonian description of quantization gives a complementary perspective which will be particularly useful for considering evolution on time slices. This is of course directly connected to the path-integral description.   For example, beginning with \fctlint, differentiate with respect to $T$ to determine the form of the evolution equation.  The quantity $T$ only enters through the asymptotic separation of the time-slices at infinity.  To be more specific, let us imagine that the amplitudes are defined in a region inside some very large radius $R$, in which case the slice separation and surface term in \ehlag\ are determined at $R$.  The time derivative of \fctlint\ with respect to $T$ then gives
\eqn\fctlsh{i{\partial\over\partial T} \calA[\Psi_i,\Psi_f,T] = \int_{\Psi_i}^{\Psi_f} \cald g \cald \phi e^{i S[g] + iS[g,\phi]}
\delta(T-\int dt N(R)) {\hat H}\ ,}
where 
\eqn\bondi{{\hat H}=-M_p^2 \int_R d^2\Omega \sqrt{{}^2g}K}
is, in the limit $R\rightarrow\infty$, the Bondi energy. 

Suppose that the asymptotic form of the background metric is $g_0=\eta+\mpm h_0$, and thus the full metric \gfluc\ behaves asymptotically as
\eqn\asymmet{g\rightarrow \eta+\mpm(h_0+h)\ .}
Then the energy \bondi\ becomes
\eqn\perten{{\hat H}=M_0 -M_p^2 \int_R d^2\Omega (\sqrt{{}^2g}K - \sqrt{{}^2g_0}K_0)=M_0+H\ ,}
where $M_0$ is the mass of the background solution\foot{Note that in the case of the semiclassical metric satisfying \backstate, this mass will actually be $T$ dependent due to the average Hawking flux.} $g_0$.  The latter term can be derived as the spatial integral of a total derivative, which is the term in the projected constraint
\eqn\const{N\calh = -\mp^2NG_{\perp\perp} + NT^\phi_{\perp\perp}}
that is linear in $h$.  From the constraint equation $\calh=0$, one then finds an expression for the hamiltonian $H$ in terms of an integral over the slices:
\eqn\hamilt{H=\int d^3x \sqrt{{}^3g_0} N_0\left(T^\phi_{\perp\perp} + t^{grav}_{\perp\perp}\right)\ ,}
where $T^\phi$ is the stress tensor for matter, and $t^{grav}$, which arises from the quadratic and higher terms in $h$ in the constraint \const, is a gravitational energy-momentum pseudotensor.  Thus, to summarize so far, the derivation of an analog of the Schrodinger equation for $T$ evolution has been outlined, with hamiltonian given by \perten\ and \hamilt.

The hamiltonian \hamilt\ generates the expected evolution of the fields on the time slices labeled by $T$.  To see this, first consider the scalar action, written in terms of the metric decomposition  \admd,
\eqn\sadmf{S[g,\phi]= \hf\int d^4x N\sqrt{{}^3g}\left[ {1\over N^2} \left(\partial_0 \phi - N^i \partial_i\phi\right)^2 - g^{ij} \partial_i\phi\partial_j\phi\right]\ .}
The canonical momentum is 
\eqn\canonp{\Pi_\phi= N {\delta S\over \delta\dot \phi} = {1\over N}\left(\partial_0\phi - N^i \partial_i\phi\right)\ ,}
and canonical commutation relations are
\eqn\ccr{[\Pi_\phi(x),\phi(y)] = -i {1\over\sqrt{{}^3g}} \delta^3(x-y)\ .}
One can likewise find the canonical momentum conjugate to the metric, which is given in terms of the extrinsic curvature $K$ of the time slices by
\eqn\metm{\Pi_{ij} ={\mp^2\over 2} (K_{ij}-g_{ij} K) \ .}
The constraint equation takes the form
\eqn\consteq{0=N\calh= 2\mpmt N\left[ \Pi_{ij}\Pi^{ij}- \hf(\Pi_i^i)^2\right] - {\mp^2\over 2}N({}^3R) + {N\over 2} \left(\Pi_\phi^2 + g^{ij}\partial_i\phi\partial_j\phi \right)\ .}
Add to this the spatial diffeomorphism constraint, $0=N^i\calh_i$, where
\eqn\spcons{\calh_i = -2D^j\Pi_{ij} +\Pi_\phi \partial_i\phi\ .}
The matter part is thus
\eqn\mattham{N\calh_\phi = {N\over 2} \left(\Pi_\phi^2 + g^{ij}\partial_i\phi\partial_j\phi + 2{N^i\over N} \Pi_\phi \partial_i\phi\right)\ .}
This precisely generates, via the commutators \ccr,  the evolution equation \canonp, and likewise the equation for $\partial_0\Pi_\phi$.  The gravitational piece $\calh_h$ of $\bar\calh=N\calh+N^i\calh_i$ similarly generates evolution of the metric perturbation $h$.

Corresponding to the expansions of the action in powers of $h$, \lagexp, \phiexp, there is an expansion of $\bar\calh$ in powers of $h$,
\eqn\hexp{\bar\calh = \sum_n \calh_n\ .}
Keeping the quadratic term in this expansion corresponds to the approximation of free quantum fields $\phi$ and $h$ in the background $g_0$, as described at the end of the preceding subsection.  Terms with $n\geq3$ describe interactions between these fluctuations.  For validity of the weak-coupling approximation, these terms should be small.

\subsec{Hawking's argument for information loss}

We are now prepared to review the argument for information loss\refs{\Hawkrad,\Hawkunc}.  Let $g_0$ describe the metric of a Schwarzschild black hole of initial large mass, $M_0\gg\mp$; we could imagine that this black hole is formed in the distant past by a collapsing distribution of $\phi$-matter, but the details of this will not concern us.  The evolution equation \backstate\ actually tells us that the geometry evolves as it radiates Hawking quanta, and in particular that its mass evolves as
\eqn\massevol{ {dM \over dT} = -\gamma {\mp^4\over M^2}}
with $\gamma$ a near-constant factor.  Since the black hole is large, curvatures are weak away from the singularity and the last instant of evaporation.  This was argued to justify use of the semiclassical approximation, which we see corresponds to keeping only $S_2$ and $\calh_2$ in our preceding description.\foot{One could clearly generalize this to describe interacting matter weakly coupled to gravity.}

Thus the state of the matter field (as well as the fluctuating metric $h$) can be derived from the viewpoint of quantum field theory in the background metric $g_0$.  Specifically, choose a set of slices that asymptote to constant Schwarzschild time slices far from the black hole, but that smoothly cross the horizon into the black hole and stay away from the singularity.  Explicit descriptions of such ``nice slices" was given by \refs{\Waldunpub,\LPSTU,\Joerev,\SGI}, and will appear in the next section.  Standard quantum field theory methods show that an initial state such as the vacuum  evolves to the form 
\eqn\state{|\Psi\rangle = \sum_{\alpha,\hat\alpha} \Psi_{\hat \alpha \alpha} |\hat\alpha\rangle |\alpha\rangle}
where the states $|\hat \alpha\rangle$, $|\alpha\rangle$ form bases for perturbations inside and outside the black hole, respectively.  
Outside observations are described by a density matrix,
\eqn\densdef{\rho = Tr_{\hat\alpha} |\Psi\rangle\langle\Psi|\ ,}
and one easily argues that this density matrix has entropy 
\eqn\Sdef{S=-Tr(\rho\ln\rho)}
which is of order the Bekenstein-Hawking value,
\eqn\bekh{S\sim S_{BH} =  M_0^2/2\mp^2\ .}

Moreover, locality/causality of quantum field theory in the background imply that the missing information described by the entropy $S$ cannot escape, at least until evolution breaks down when the black hole approaches the Planck size.  The simplest possibility is that the black hole completely and rapidly evaporates once it reaches this regime.  But in this case, basic considerations of energy and entropy imply that the missing information cannot immediately escape the black hole.  Indeed, as argued by Page\refs{\Page}, if information is to escape, it must begin to do so by the time the black hole has appreciably evaporated, say to size $M_0/2$, and thus on time scales $T_{evap}\sim M_0^3$.  As a consequence, information was argued to be destroyed, violating unitary quantum evolution\Hawkunc.  Another alternative is that a black hole leaves behind a long-lived remnant containing the information, but such scenarios appear ruled out by their prediction of infinite remnant species and thus divergent production rates.

This scenario produces  a paradox, since an effective description of such information loss was argued\refs{\EHNS,\BPS} to lead to violent conflict with energy conservation, corresponding to effective vacuum temperatures of order $\mp$. 

\newsec{Quantization on nice slices}

To improve our understanding of  the information loss argument, let us examine it in the context of our description of quantum gravity and the emergence of the weak-coupling/semiclassical approximation.

\subsec{Nice slices and semiclassical states}

Begin with a more precise characterization of nice slices.  As stated in the previous section, these should asymptote to slices of constant Schwarzschild time, but also enter the horizon while avoiding the singularity.  Such a choice of slicing corresponds to a particular choice of gauge.  There are many ways to choose such slices, but they all share certain characteristics.  In particular, remaining spatial while avoiding the strong curvature regime requires that inside the black hole they be closely spaced together in the large $T$ limit; in other words, the lapse $N$ becomes extremely small, roughly of order $1/T$.  There are various ways that one could choose the bound on how close such slices get to the singularity;  in the semiclassical approximation one could specify an explicit constraint on the minimum radius $r_c$, or alternatively on the maximal tetrad components of the curvature, maximal extrinsic curvature of the slice, {\it etc.}.  When reformulated in the quantum framework, all of these correspond to different choices of gauge, specifically the surface conditions described below eq.~\resid.

Let us give an explicit example.  One family of nice slices that is particular convenient for our purposes can be constructed by taking a constant Schwarzschild time slice down to some outside radius $r_o\sim 10M_0$.  At this point, we attach an arc, with moderate extrinsic curvature, that smoothly joins the $r>r_0$ piece of the slice to the curve $r=r_c$ inside the horizon.  Once these curves meet, the rest of the nice slice, at Schwarzschild $t$ coordinate before the joining point, is simply taken to be the surface $r=r_c$, where, say, $r_c\sim M_0/10$.   The family of slices is generated by acting on this slice by a Schwarzschild time translation.\foot{More carefully, the geometry $g_0$ does not have an exact Killing vector; so one duplicates this construction for later Schwarzschild times, requiring a small deformation in the arc piece for large times.  This will not be critical to our arguments.}  In particular, note that in this gauge the lapse $N$ vanishes where the slices degenerate at $r=r_c$.  Evolution is frozen here.  This makes sense from the perspective of the scalar lagrangian, \sadmf:  fluctuations of $\partial_0\phi$ around $N^i\partial_i\phi$ are suppressed -- the only possible evolution is due to a non-trivial shift.

 \Ifig{\Fig\fignice}{A sketch of two nice slices in the Eddington-Finkelstein representation of the black hole geometry.  The dotted line is the horizon.  The stretching effect is easily seen, as the distance from point $P$ to radius $r_o$ on subsequent slices increases linearly with $T$. }{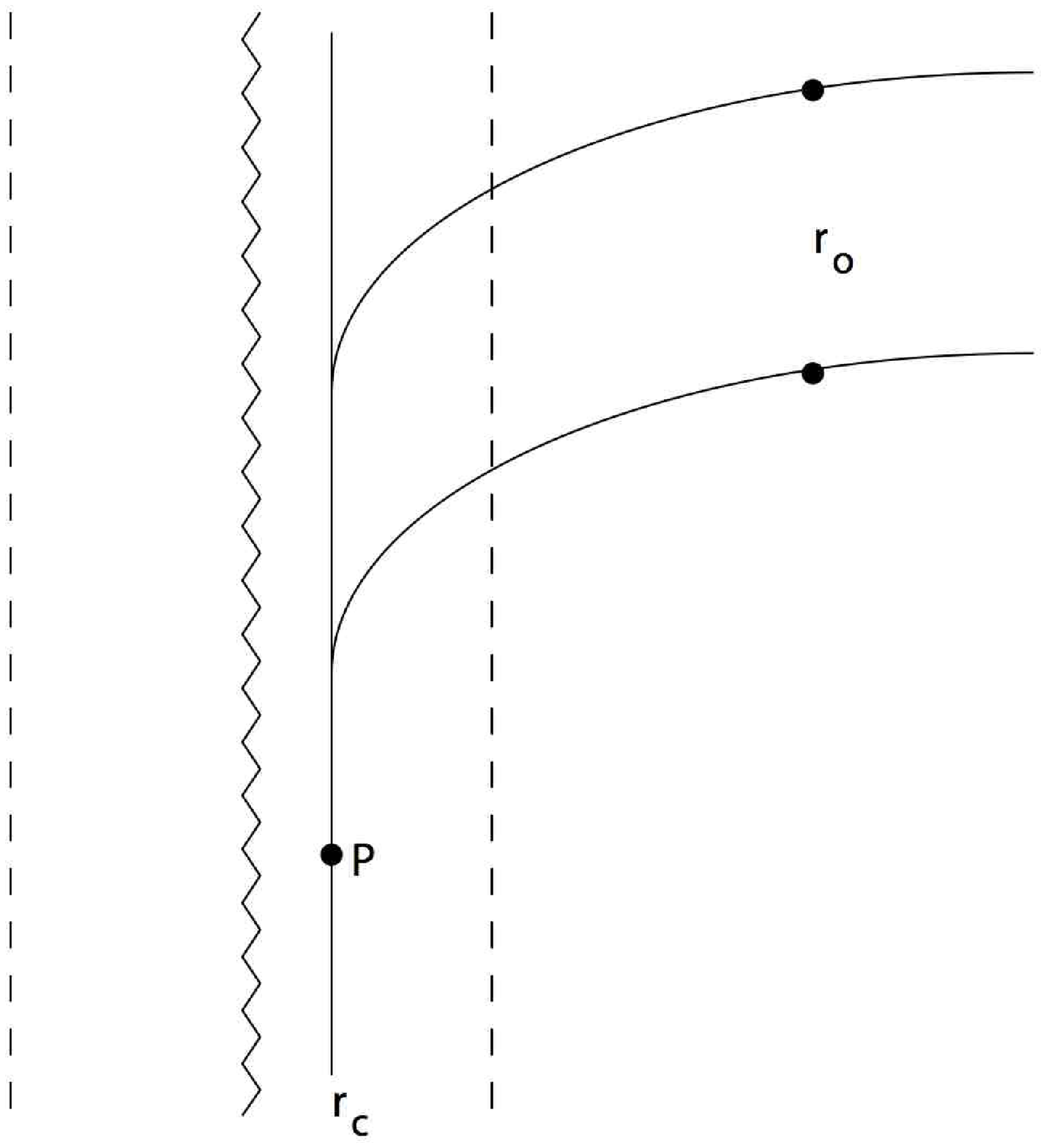}{3}

An interesting feature of such nice slices is that  they continuously grow by stretching.\foot{For one description of this phenomenon and an attempt to connect it to breakdown of the semiclassical approximation, see \refs{\Mathur}.}  Specifically, consider a point $P$ on the $r_c$ section of the $T$ nice slice, and compute the distance in the  intrinsic slice geometry
to the point at radius $r_o$.  Then consider a later slice, with time $T+\Delta T$.   The distance from the point $P$ to $r_o$ along this slice is larger by an amount
\eqn\stretch{\Delta s = \Delta T\sqrt{{2M\over r_c} -1}\ .}
This arises because an extra piece has been added to the slice at $r=r_c$; the rest of the slice is the same.
While the precise formula \stretch\ is particular to our choice of slices, the generic phenomena holds for other choices, perhaps with more complicated formulas for the stretch.

Next consider the state on the nice slice that is predicted by Hawking's arguments\Hawkrad\ and subsequent refinements.  In the semiclassical approximation, quanta are produced in pairs near the horizon.  These have a thermal distribution of asymptotic energies and wavelengths at the Hawking temperature 
\eqn\thawk{T_H={\mp^2\over M}\ .}
One can work in a basis where the asymptotic quanta are wavepackets, which when described in terms of asymptotic retarded time $u$ have widths $\Delta u\approx M$.  A corresponding basis can be chosen for the paired quanta inside the black hole.  Specifically, if we consider the initial state of the $\phi$ field to be the (in) vacuum, we have
\eqn\vacexp{|0\rangle = \sum_{\{n_k\}} e^{-E(\{n_k\})/2T_H} |\{\hat n_k\},\{n_k\}\rangle}
where  $\{\hat n_k\}$  and $\{n_k\}$ label the states by occupation numbers for the inside and outside wavepacket bases.  (For more details, see \refs{\SGI}.)  One can think of a ``typical" state as a state where one Hawking quantum of wavelength $M$ is emitted from the black hole for each (asymptotic) time step $\delta T=M$; this indeed reproduces the evolution law \massevol.   Correspondingly, the inside partner quanta (with paired correlations described by the state \vacexp) evolve towards $r=0$ and in the nice-slice description freeze at $r=r_c$.  For asymptotic time-step $\delta T$, the corresponding proper-distance separation between the ``impact points" of the centers of the inside wavepackets on the nice slice is 
\eqn\nicedist{\delta s = \delta T\sqrt{{2M\over r_c} -1}\ ,}
and this formula also gives their characteristic proper-distance width.  So, in summary, the Hawking state consists of a sum over states with paired inside and outside quanta, where the inside and outside quanta are spaced by typical proper distances along the slice $\delta s$ and $\delta T$ respectively.

\subsec{Perturbations in nice-slice gauge}

As explained in both sections 2.1 and 2.2, the approximation of quantum field theory in a semiclassical background can be derived by taking the leading order terms in an expansion in $\mp^{-1}h$.  Specifically, one drops the terms $S_3$ and higher in \lagexp, \phiexp, or $\calh_3$ and higher in \hexp, that describe couplings of the fluctuations in $\phi$ and $h$ to $h$.  Moreover, as argued in section 2.3, this is precisely the approximation in which a refined version of Hawking's argument for information loss is made.

In examining the validity of the argument for information loss, one would like to assess the validity of such a weak coupling expansion.  In particular, this means examining the relative size of the subleading terms in the expansion \hexp.  For a measure of this, phrase the question in the context of usual perturbation theory.  If we have a quantum-mechanical system with hamiltonian $H=H_0+H'$, where $H'$ is considered a perturbation of $H_0$, the effect of the perturbation can be easily described in interaction picture.  The time-dependent perturbed state takes the form
\eqn\pertstate{|\psi(t)\rangle = T\exp\left\{ -i \int^t dt H'(t)\right\} |\psi\rangle_0, }
where $|\psi\rangle_0$ is the unperturbed state.  Perturbation theory is valid as long as matrix elements of $\int dt H'(t)$ are small.  In contrast, suppose that 
\eqn\orthocon{\langle\psi(t)|\psi\rangle_0=0}
 for some $t$.  This would be an indicator that perturbation theory has broken down, since in this case $\langle\delta\psi|\psi\rangle_0 =-1$: the perturbation has effect of order unity.  

Specifically, in the present context, think of the terms $S_2$ and $\calh_2$ as describing the unperturbed evolution, and the terms $S_n$ and $\calh_n$ for $n\geq3$ as describing a perturbation of the hamiltonian.  One would like to assess whether these higher-order terms indeed have small effect on the evolution, for example as measured by their effect on the state as described in the preceding paragraph.

Thus we will consider the difference in evolution and in particular in final nice-slice states with and without the $n\geq3$ terms.  In order to most cleanly exhibit the effect, consider an initial state that is a small perturbation of the ``vacuum" black hole state.  Specifically, one could consider a perturbation that corresponds to throwing one quantum into the black hole, at some early time, designated $T=0$, changing its mass by $m$.  A minimal such change is 
\eqn\msize{m\sim \mp^2/M\ ,}
 corresponding to the energy of one Hawking quantum.  Or, if we think of the black hole evaporation as a statistical process, there will be fluctuations.  The average energy flux is given by \backstate, but we might consider a state where there is a (typical) fluctuation such that one expected Hawking quantum fails to be emitted.  This can be made explicit by for example projecting the state \vacexp\ onto states with a ``missing" quantum in a particular mode $k$ corresponding to an outgoing Hawking particle at time $T=0$.  This will change the mass of the hole by $m$, also given by \msize.  Such a fluctuation is representative of a typical fluctuation of the full quantum dynamics.  Our goal is to compute the difference in the resulting state at some later time $T$, in either of these cases, that arises from including or excluding the $n\geq3$ terms in the Hamiltonian.
 
 Of course, the approximation where $n\geq3$ terms are dropped is precisely the approximation where the backreaction of the fluctuation on the metric is neglected; in this case the state $|\Psi(T)\rangle$ is simply given by the original Hawking calculation, as described above.
 
 One can estimate the effects of the perturbation terms by considering a perturbation that is purely s-wave, as is in fact typical of the Hawking quanta.  To compute the resulting state, one can shift the background $g_0$ to include the effect of the perturbation.  For example for an infalling quantum of mass $m$, this would be a new metric that is approximately described as an ingoing Vaidya solution\refs{\Vaid}, together with the effects of Hawking radiation described as in \backstate.  Or, for a perturbation such that one Hawking quantum fails to escape, this would be a solution that looks like outgoing Vaidya, with the source stress tensor \tno\ describing a positive energy particle falling into the singularity, and the outside fluctuation represented by  a negative energy particle travelling out to infinity, raising the mass of the hole by $m$ at $T=0$, with subsequent evolution again given by \backstate.
 
For simplicity, in evaluating the effect of the perturbation $m$ we will ignore the Hawking flux represented by the stress tensor in \backstate, although analogous statements can be argued to hold there as well.  Thus, we treat the perturbed geometry as simply Schwarzschild with mass $M_0+m$.

How is the nice-slice state changed?  This of course depends on choice of slicing, {\it i.e.} gauge.  In other words, clearly one can take a slicing of the perturbed spacetime such that there {\it appears} to be a big effect.  To avoid this, one tries to specify gauge conditions that are ``as nice as possible," in the sense that they lead to {\it minimal} change in the resulting state.  Outside the black hole, one can in this spirit take the new nice slices to still be described as Schwarzschild time slices far from the black hole, and take some small modification of the arc crossing the horizon.  Inside the black hole, there are a number of options.  The simplest is to keep the nice slices at the radius $r_c$ determined by the original geometry; in this case the angular part of the intrinsic geometry of the slices is fixed.  Alternately, one might choose a gauge such that $\nabla^i {\bar h}_{ij} =0$ on the slices, or $\delta K_{ij}=0$, or even such that components of the Riemann tensor are fixed on the slices.  Any of these conditions, and a multitude of others, will lead to a fractional change of $M/r_c$ of order $m/M\roughly>\mp^2/M^2$.  In particular, this means a small change of the relation \stretch\ for stretch versus asymptotic  time,
\eqn\stretchp{\Delta s = \Delta T \sqrt{{2M\over r_c} -1} [1 + \calo(m/M)]\ .}
Likewise the description of the evaporation changes by $\calo(m/M)$  in the Hawking temperature, and thus energy of the typical outgoing Hawking quanta on the nice slices, as well as in the spacing $\delta s$ of the wavepacket crests hitting the nice slice, and in their widths.

We might compare the resulting perturbed nice-slice state $|\Psi'(T)\rangle$ to the state where we ignore the effects of $\calh_3$, {\it etc.}, by taking their inner products.  Considering for example the $r=r_c$ part of the slice, they both have corresponding wavepackets that hit the slices at proper distance intervals separated by $\delta s$ or $\delta s'$, which are nearly identical.  Thus the contribution to the inner product from corresponding wavepackets is essentially unity, to $\calo(m^2/M^2)$.  Quantum-by-quantum, the states are nearly identical.

However, if we take into account the contributions of quanta corresponding to Hawking emission at very late times, this changes.  For a first estimate, note that the cumulative shift in the center of the $n$th quantum in the perturbed state, relative to the unperturbed state, is of order 
\eqn\cumshift{ s_n'- s_n \sim n {m\over \mp^2} \sqrt{{2M\over r_c} -1}\ .}
A crude estimate is that when this becomes of order the size of the wavepackets, \nicedist, then the overlaps of the wavepackets, and thus the states, nearly vanishes.  This occurs for a time $T$ given by 
\eqn\tbreak{T\sim {nM\over \mp^2} \sim {M^2\over m \mp^2}\ ,}
or, in a $D$-dimensional generalization,
\eqn\tbreakd{T\sim \left({M\over \mp}\right)^{D-2\over D-3}{1\over m}\ .}
For an initial perturbation of size \msize, this is
\eqn\tmc{T\sim M^3/\mp^4\ ,}
with generalization
\eqn\dbreakd{
T\sim \mp^{-1} \left({M\over\mp}\right)^{D-1\over D-3}\ .}
For a larger perturbation $m$, it occurs even earlier.

\Ifig{\Fig\statecomp}{An illustration of the comparison of the state on the internal part of a nice slice with and without the effects of the perturbations $\calh_n$, $n\geq3$.  Vertical marks represent the centers of corresponding wavepackets.  Initially the mismatch is small, but it increases along the slice.  A similar mismatch develops on part of the slice outside the black hole.}{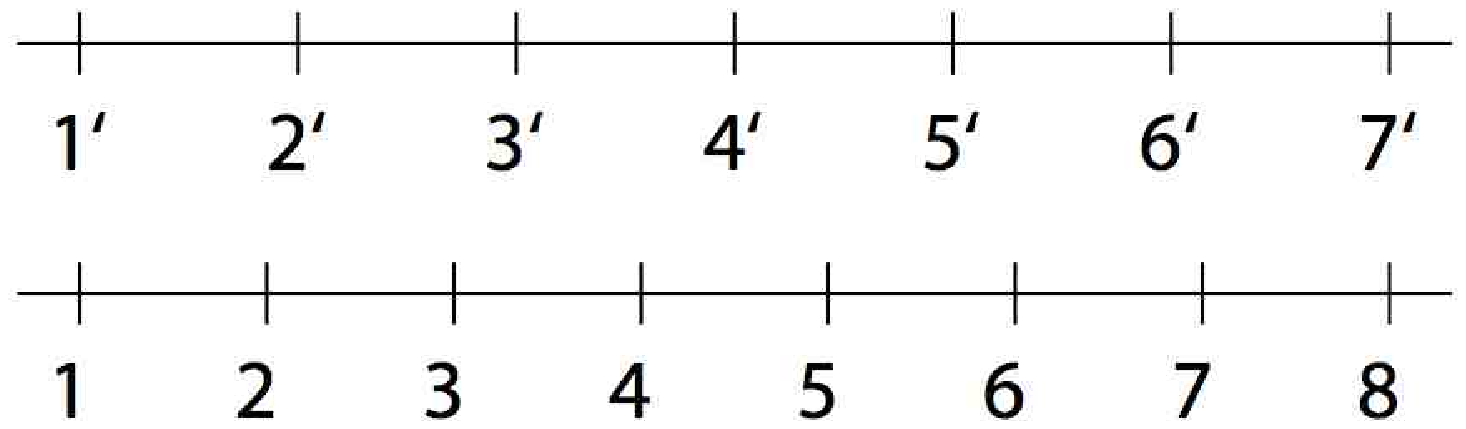}{4}

Therefore, by the time \tmc\ the  perturbations $\sum_{n\geq3}\calh_n$ have an effect that is of order one, by a criterion analogous to \orthocon; the same effect also deforms the state outside the black hole.    This effect indicates that this perturbative expansion is not valid at such times.  One way of thinking of this is that the shift in the hamiltonian is very small.  However, over long times it has a coherent effect on the state, and thus can have a significant net effect.  This is a failure of the leading order treatment of the problem in the weak-coupling expansion.  As a result, the leading order $n=2$ perturbative treatment, which we've seen produces the 
 semiclassical description of \Hawkrad, does not accurately compute the state at times given by \tbreak, or the corresponding density matrix.

Indeed, note that the {\it net} effect on the nice slice state becomes important even before the time \tmc.  Specifically, given the offset \cumshift\ and 
for gaussian wavepackets of width $M$, the inner product between the $nth$ perturbed wavepacket and the $nth$ unperturbed wavepacket is thus of order
\eqn\packover{{\cal P}_n \approx e^{-(nm/M)^2}\ .}
The inner product between perturbed and unperturbed states, accounting for the first $N$ quanta, is thus
estimated as
\eqn\innp{{\cal P}\approx \prod_{n=1}^N {\cal P}_n \sim e^{-N^3(m/M)^2}\ .}
This product is small past the time
\eqn\smtime{T\sim {M\over \mp^2} N \sim {M^{5/3}\over m^{2/3} \mp^2}\ ,}
or $T\sim M^{7/3}$ in Planck units, for $m$ given by \msize.  The time estimate \tbreak\ is an estimate for where the effect on {\it each quantum} becomes significant, but before this, at the time \smtime, the net effect on the state becomes significant.  (Notice that the argument of \refs{\Page} only tells us {\it by} what time scale there must be some interactions that relay information if information is to emerge in the Hawking radiation.)

\newsec{Discussion}

\subsec{Diagrammatics}

To better understand the result of the preceding section, begin by describing it in terms of Feynman diagrammatics, arising from expanding the functional integral \ampgff, or equivalently the evolution operator derived from the hamiltonian \hexp, order-by-order in  $1/\mp$ about the quadratic approximation, so perturbatively in the terms $n\geq3$ of the action or hamiltonian. 

Notice that in the black hole context, there is a generalization of loop diagrams.  In particular, simply in the quadratic approximation, one can have a nonzero amplitude of the form
\eqn\loopgen{\langle \Psi_f| \Delta T_{\mu\nu} |0\rangle\ }
where the operator $\Delta T_{\mu\nu}$, which could be the energy-momentum tensor either for $\phi$ or for $h$ fluctuations, is normal-ordered as in \tno\ with respect to an in-basis, and thus can create a non-trivial out state with an extra pair of particles: one escapes to infinity, and the other falls into the strong-curvature regime.  

\Ifig{\Fig\feyndiag}{A Feynman diagram representation of the coupling between fluctuations at different times.  The wavy line is the graviton propagator.}{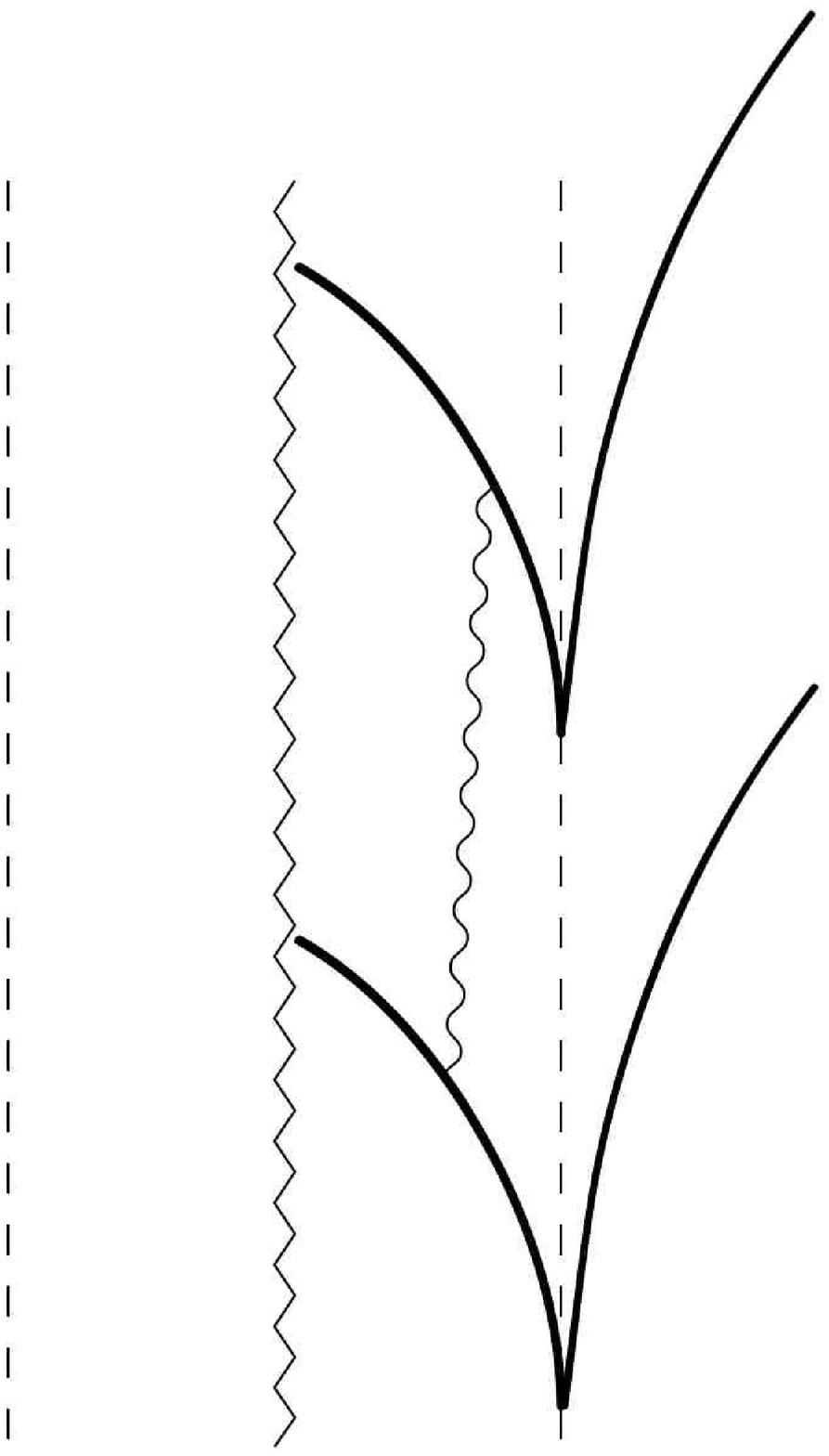}{2.5}

When one includes the cubic and higher interactions, one also has diagrams with such ``open loops" connected by graviton propagators.  For example, the second-order term in the expansion of $\exp\{iS_3\}$ has a contribution corresponding to a pair of vacuum fluctuations connected by a graviton propagator.  The case where the vacuum fluctuations are separated by a large time $T$ is the case described in the preceding section.  Specifically, the effect of the first fluctuation on the metric was computed, and then this in turn interacted with the second fluctuation to produce a state that deviates from the state with no interaction terms.  This amplitude could equally be represented by a Feynman diagram, of the form
\eqn\feynrep{\langle\Psi_f'| \Delta T_{\mu\nu} D^{\mu\nu,\lambda\sigma}\Delta T_{\lambda\sigma} |0\rangle}
where we consider states $\Psi_f'$ on the late-time nice slice that contain both the early and late fluctuations.  Here $D^{\mu\nu,\lambda\sigma}$ is the graviton propagator.  The large phase shift described in the preceding section corresponds to the amplitude \feynrep\ giving a large correction.  Likewise, one could consider higher order amplitudes, {\it e.g.} expanding to higher order in $S_3$ or in higher terms in the action.

\subsec{Large boosts}

From this perspective an obvious question is the origin of the large effect.  One way of thinking of this is to note that for large time separations $T$, there is a large relative boost between the two fluctuations.  This boost can be estimated by integrating the extrinsic curvature along the nice slice.  Specifically, the change in the normal to the slice under a small displacement $dl$ is 
\eqn\deltan{d n_i= K_{ij} dl^j\ .}
Since the extrinsic curvature takes the constant value 
\eqn\kcons{K_{tt}= -{M\over r_c^2}\sqrt{{2M\over r_c}-1}}
along the slice, the net boost parameter between two points separated by $T$ is 
\eqn\netboost{\theta=\int E^t K_{tt} dt = {M\over r_c^2} T}
where $E$ is the unit vector in the $t$ direction.  Thus the boost is exponentially large in the time separation $T$.  This was argued in \refs{\GiLione,\SGI,\SGIII,\BaFi}  to explain how small early fluctuations could have large effects on the late metric and late fluctuations, since large boosts magnify gravitational fields.  In the preceding section, we have indeed argued that the effect on the late-time state  is large.  While this is apparently related to the large boost, the relevant
calculation of the graviton propagator in nice-slice gauge yields an effect on the state that only grows as $T$.  In any case, the significant coupling between perturbations appears closely parallel to the origin of strongly coupled gravity in a high-energy collision in a flat background\refs{\SGII}.

\subsec{Lessons, and nonlocality}

The results of this paper suggest the following possible conclusions.  First of all, the present weak-coupling (or semiclassical) expansion of quantum gravity, which we have shown produces the approximation of quantum field theory in a curved background, apparently breaks down for the purposes of calculating late-time states on nice-slices.  Validity of the semiclassical approximation has been assumed in \Hawkrad\ and in subsequent calculations of the state.  It may be that there is a different way of constructing a perturbative expansion that gives a controlled calculation of the state, but this author does not see the concrete evidence for this at present.

This approximation may certainly yield valid results for some quantities, like the expectation value of the stress tensor outside the black hole, {\it etc.}, as such quantities can be computed without explicit reference to the state on the nice-slice.  However, a calculation critical to  the information paradox is that of the density matrix describing the outside world, and its entropy.  Existing calculations of this rest heavily on calculating the combined state inside and outside the black hole, and tracing over the internal state, and the most precise extant versions of these apparently rest on logic that amounts to using nice-slice states.  For the purposes of computing the density matrix and its entropy, one must do a very fine-grain and precise calculation.  In the present approach to this calculation we have exhibited an apparent failure of the semiclassical approximation.

If the state and density matrix cannot be reliably calculated, this calls into serious question the arguments given for an information paradox -- and if reliance on an unreliable calculation produces a paradox, the conclusion appears obvious.  However, one would like to go further and understand how one could perform a reliable calculation, and how such a calculation would demonstrate information escape from the black hole.  The preceding discussion is consistent with the suggestion that such a calculation must  be given in the context of a non-perturbative description of quantum gravity, since the perturbative approach appears to fail over {\it macroscopic} distances.  

In particular, it {\it could} be that there is a non-perturbative calculation of the precise analog of a nice slice state and outside density matrix that retains the essential features of local quantum field theory, and in particular implies that the nice slice state has information content deep inside the black hole that is not revealed to the outside world.  This author regards this assumption about the non-perturbative dynamics as in doubt, as it returns us to the original paradox.  

Thus one might consider other alternative assumptions about non-perturbative gravitational dynamics.  There are several reasons to question the assumption of locality.  First, there are no known precisely local observables in a quantum theory of gravity; the best one can apparently do is to construct certain relational quasi-local observables that approximately reduce to local observables in certain states\refs{\GMH,\GaGi}.  This suggests that there is no precise notion of locality in quantum gravity.  Second, if one attempts to probe locality via the S-matrix, in high-energy scattering, one finds characteristic growth of the cross section with energy $\propto E^2$ in four-dimensions, which violates expected bounds on local theories. Third, recent developments in string theory, and in particular the AdS/CFT correspondence\refs{\Mald}, suggest that there may be a complete ``holographic" boundary description of dynamics in a spacetime region, in contrast to the tenets of local field theory.  Finally, there is the paradox itself, which seems inevitable if physics is local. 

In fact, the effects described in the preceding section, while indicating  a breakdown of reliable calculations, appear to hint at a nonlocal effect, as the fluctuations on the nice slice deep within the black hole develop unanticipated correlations with the fluctuations in the vicinity of the horizon.  And, as described above, the origin of the effect appears closely connected with the dynamics of high-energy gravitational scattering.   

Thus, these arguments suggest a possible entree for an alternative assumption about nonperturbative gravitational dynamics, that of a {\it nonlocality principle}\refs{\SGI,\GiLione,\GiLitwo,\SGII,\SGIII}, stating that gravitational dynamics is indeed fundamentally quantum mechanical, but that the dynamics that unitarizes gravity at strong coupling does not have the usual locality properties of local quantum field theory.  Criteria for operation of such a nonlocality principle were suggested in \refs{\SGIII,\GiLione,\GiLitwo\SGI}; in a flat background these were phrased in terms of a locality bound (with n-particle generalization in \SGII), stating that when two modes with sufficiently high center-of-mass energy ({\it i.e.} relative boost) have sufficiently small separation, a description via local QFT fails.  As we've seen above, the present criterion for breakdown of the semiclassical approximation appears closely related, and in particular breakdown of the perturbative expansion arising in
the black hole geometry apparently prevents a straightforward semiclassical treatment.

One can guess at how such nonlocal dynamics could yield unitary evolution.  Specifically, a non trivial state which initially appears of the form
\eqn\initst{|\psi_i\rangle= \sum_{\alpha,\hat \alpha} \psi_{\alpha\hat\alpha} |\hat \alpha\rangle_{in}|\alpha \rangle_{out}\ ,}
and thus contains entanglement between inside and outside degrees of freedom, might by such nonlocal dynamics evolve to a state of the form
\eqn\finstate{|\psi_f\rangle = |BH\rangle_{in}|\psi_f\rangle_{out}\ }
as described from the perspective of the outside observer, where the state $|BH\rangle$ is an internal state that doesn't contain information about the initial state.  This is essentially the final state proposal of \refs{\HoMa}.  What was lacking in that proposal was an explanation of how local physics could be violated to give such evolution;  a breakdown of the semiclassical approximation together with the proposed nonlocality principle would provide such a rationale.\foot{For one discussion of the nonlocality of the final state proposal, see \AhnZK.}

As stated, a reasonable viewpoint is that the black hole information paradox is a challenge to local quantum field theory of the same magnitude of the challenges to classical physics that ultimately catalyzed the invention of quantum mechanics.  Specifically, the classical instability of matter serves as a sharp example of what is wrong with the classical model of the atom.  While one could imagine a na\"\i ve 
classical physicist approaching this as a mathematical problem to be resolved by some modification of the singularity
in classical evolution where the electron reaches the charge center, we know that this approach is far from the mark:
the classical description fails to be the correct description, and must be {\it replaced} by a quantum description, at scales set by the Bohr radius.  Classical physics is invalid at the Bohr radius, without there being any explicit signal {\it within} the classical theory that it fails there.  

Likewise, many gravity theorists have been perplexed about the dynamics at a black hole singularity, and have made various proposals about what smooths singularities.  But the information paradox suggests that local quantum theory could break down on scales of order the Schwarzschild radius.  This could plausibly be the ultimately correct picture, whether or not we had found a signal within existing theory  in the form of an explicit breakdown of local QFT at the Schwarzschild radius.  The arguments of the preceding section indeed suggest that history has  not repeated itself precisely:  they are consistent with an actual breakdown of a local quantum analysis.   But whether or not this is an ultimate indicator of new nonlocal quantum physics, such nonlocal physics could be the correct description of quantum dynamics on Schwarzschild scales, just as quantum dynamics replaces classical dynamics at atomic scales.

Independent of the preceding arguments, there is another argument suggesting that one question the argument for information loss.  This is Wheeler's paraphrase of Bohr's dictum: ``no phenomenon is a real phenomenon until it is an observed phenomenon."  The information loss argument relies heavily on the notion of the internal state on a nice slice.  But one must be careful in specifying a gauge-invariant description of this internal state.
Consider what would be needed to actually measure this state, in the spirit of the discussion of quasi-local observables of \refs{\GMH,\GaGi}.  In essence, the state can only be measured by constructing detectors that it correlates with.  So, one could construct detectors at infinity, and build them with sensors that activate measurements when they reach the radius corresponding to the nice slice.  A family of such ``depth-charge" detectors thus in principle could apparently determine the state on the nice slice.  However, to do so accurately one would have to make a measurement for each spatial interval of size $\delta s$ given in \nicedist.  An absolute (and unrealistically low) minimum for the energy each  detector would have is estimated as $\calo(\mp^2/M)$.  A family of such infalling detectors would thus have a huge effect on the black hole -- and in particular it would deviate from the undisturbed black hole on the $M^3$ time scale, by failing to evaporate.  Notice that these ideas that we must be able to measure the state actually involve a weaker assumption than those of black hole complementarity\refs{\BHcomp}, in that the latter discussion describes both local measurements and moreover requires that the results be relayed to a {\it common observer}.  

\subsec{Inflationary cosmology}

One anticipates similar effects might play a role in breakdown of perturbative local effective theory in other spacetimes with horizons, and in particular in inflationary cosmologies.  This is suggested by the following rough argument.  Consider de Sitter space.  Here one can think of the spheres of the global description as ``nice slices."  Thermal de Sitter radiation corresponds to the de Sitter horizon emitting roughly one quantum of wavelength $\lambda\sim R_{dS}$ in each efolding time $R_{dS}$.  

 \Ifig{\Fig\figds}{An extended Penrose diagram for de Sitter space.  Horizontal lines correspond to $D-1$ spheres which are thought of as nice slices; $N$ and $S$ label the north and south poles, respectively.  A fluctuation that is $s$-wave with respect to the observer at $N$ produces a pair of particles (shown as dashed lines) on either side of the horizon.  Regions I and III are thus de Sitter space, but region II is described as Schwarzschild-de Sitter.  The observer at $O$ is as a result expected to see a significant late-time phase shift in the nice-slice state, relative to the corresponding state without backreaction included.}{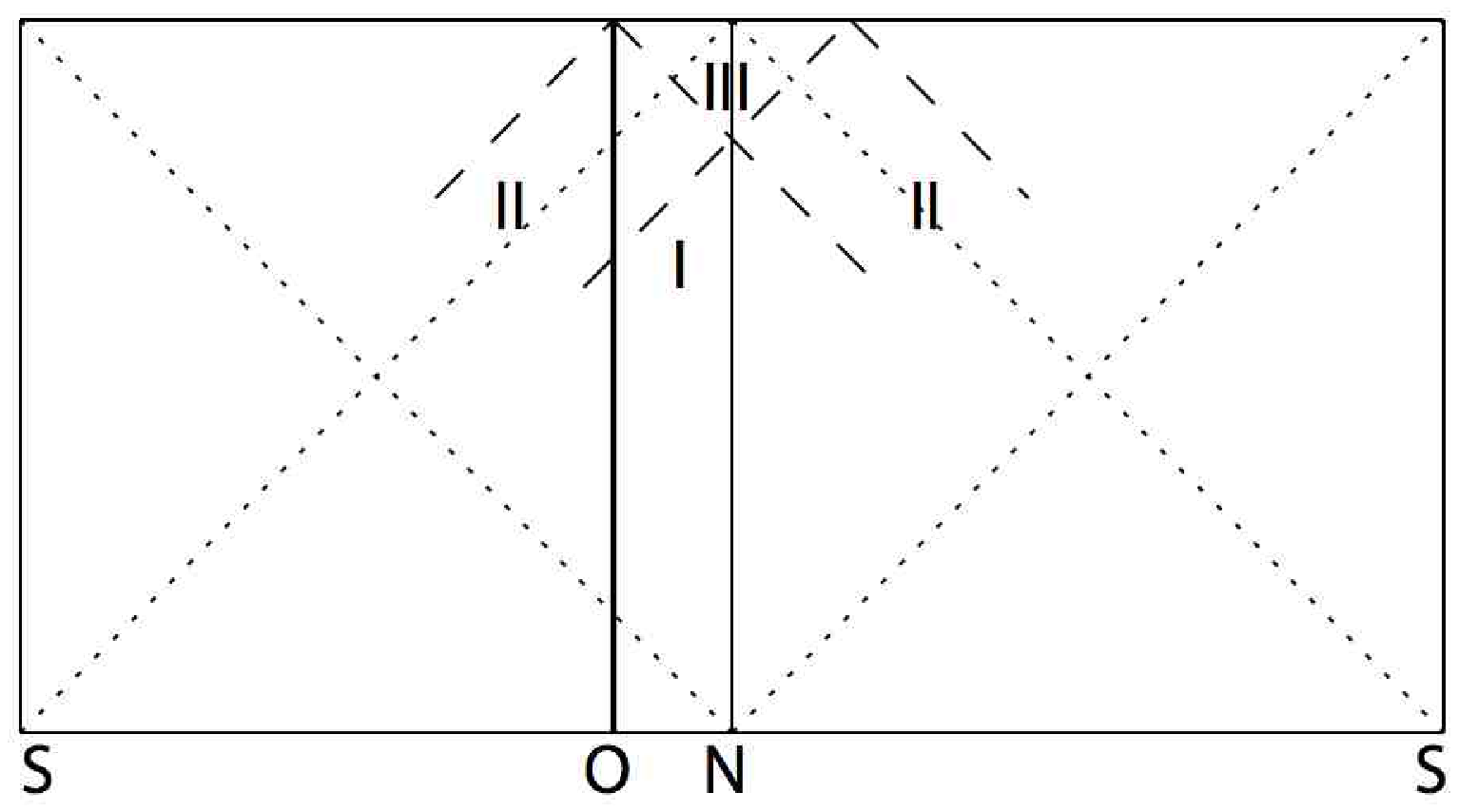}{5}

For simplicity, focus on the $s$ wave sector with respect to a definite observer taken by convention to reside at the north pole.  In analogy with the black hole, pairs of quanta are produced on either side of this observer's horizon.  These follow the trajectories shown in \figds.  Regions I and III are de Sitter, but region II should be Schwarzschild-de Sitter, resulting from the energy perturbation.  The observer $O$ thus appears to experience late evolution in this Schwarzschild-de Sitter region.  This observer is expected to see a thermal spectrum with a temperature shifted by the fractional amount $\delta m/R_{dS}$ where as in \msize\ we expect the minimum such change given by $\delta m\sim 1/R_{dS}$.  Correspondingly, after a time $T$, the state on the nice slice is expected to experience a net shift in the positions of the quanta analogous to \cumshift.  Thus, for $T\sim R_{dS}^3$, the shifts are estimated to become of order $R_{dS}$.  This would suggest that the perturbative expansion producing local field theory breaks down after $N\sim S_{dS}\sim R_{dS}^2$ efoldings, and at this time an original region of size $R_{dS}$ has inflated to size $\sim R_{dS} e^{S_{dS}}$.  For later times/larger volumes, this argument suggests a perturbative local description is not valid.  Similar statements have been inferred from the general analogy between black hole and de Sitter horizons by Arkani-Hamed\refs{\NAH}.  There may be other manifestations of breakdown of local field theory in de Sitter space as well; some previous work on this subject includes\refs{\Bankslittle\Fischler\BoussoFQ\DysonPF\BanksWR\GoheerVF-\BanksCG} and these issues are presently under investigation\refs{\GiMa}. 

\newsec{Conclusion}

This paper has described an attempt to carefully justify the derivation of the state on a set of nice slices through a black hole geometry, in a semiclassical approximation like that used in \Hawkrad, and in subsequent improvements, to argue for fundamental information loss.  While one can in principle derive local quantum field theory in a leading order approximation, in the limit $\mp\rightarrow\infty$, to either the path integral or to a corresponding hamiltonian evolution,  there are apparently circumstances where such a perturbative approach fails.  One example is that of high-energy scattering\refs{\SGII,\SGIII}, and this paper has presented arguments apparently consistent with the statement that
 perturbative gravity fails to give accurate results  in computing precise states on nice slices for times $T\roughly>\calo(M^3)$.  Thus the present approach to perturbative gravity does not appear to accurately give the density matrix or its entropy.  This happens by a time scale of order the the time by which information would need to escape the black hole\refs{\Page} if evolution is indeed unitary.

Since assuming a valid perturbative calculation apparently leads to the information loss paradox, an invalidity offers a welcome possible out.  Specifically, if there is no reliable calculation of the information destroyed by a black hole, there is no information paradox.  Beyond this, to actually explain how information escapes a black hole, one would apparently require some new dynamics of an essentially nonlocal character.  There are several reasons to question the essential nature of locality in non-perturbative quantum gravity:  rapid growth of high-energy scattering amplitudes, fundamental limitations on local observation\refs{\GMH,\GaGi}, suggestions from string theory via the AdS/CFT correspondence, and finally the magnitude of the paradox itself.  Thus, the combined evidence plausibly indicates the existence of a gravitational nonlocality principle, stating that  
 the nonperturbative dynamics of gravity has a certain non-local quality, or alternatively, that the  usual notion of locality is not an essential part of the theory and is only recovered in an approximation.  If this is correct, a key question is:  what is this dynamics?

\bigskip\bigskip\centerline{{\bf Acknowledgments}}\nobreak

The author wishes to thank N. Arkani-Hamed,
 G. Horowitz,  D. Marolf, S. Mathur, and especially J. Hartle for important discussions.
This work  was supported in part by the Department of Energy under Contract DE-FG02-91ER40618,  and by grant 
RFPI-06-18 from the Foundational Questions Institute (fqxi.org).


\listrefs
\end